# Quantum Observables and Ockham's Razor

J.M. Picone*

Space Science Division

Naval Research Laboratory

Washington, DC 20375

July 28, 2022

* Emeritus, j.picone.ctr@nrl.navy.mil or jmpicone@verizon.net

## Abstract

For the paradigm of the quantum double-slit experiment (DSE), we apply Ockham's Razor to interpret quantum observations and to evaluate terminology often associated with wave-particle duality. One finds that the Correspondence Principle (CP), combined with classical wave DSEs, e.g., Young [1804], is sufficient to separate and anticipate the observed quantum particle and wave phenomena. The empirical approach of Ockham infers that individual quanta are only whole particles during transit from source to detector; an individual quantum never acts as a wave. The wave nature of quanta emerges only in the distribution of large numbers of single-quantum observation events. That is, the "measurement problem" is no problem at all; "particle" and "wave" derive from separate and different aspects of a set of observations. Such artificial constructs as wave function collapse are irrelevant to the observation of individual quanta, each of which acts as a whole "particle" from emission to measurement. The histogram of detected events is a collective property identical to a classical interference pattern in the limit of large numbers, as the CP decrees. For a specific individual quantum, Ockham's razor renders irrelevant any attributed mystery or hypothesis regarding wave- or particle-like behavior of the quantum in the region between emission and detection. This region has so far been inaccessible to direct observation in coincidence with detection of the same quantum on the DSE screen, and such speculation is therefore inessential to interpreting the extant database. To analyze actual data sets consisting of a large number of identical observation events and to predict future DSEs, therefore, the Correspondence Principle of standard quantum (wave) mechanics is sufficient: in the limit, the distribution of observations approaches the continuous density defined by the standard quantum mechanical wave function. True scientific progress beyond this picture requires new, relevant experiments. From the DSE, Ockham's Razor infers that a theoretical quantum system consists of at least one quantum particle plus a wave function specifying the statistical distribution and properties of a large number of identical such particles.

1. Introduction: "Wave-Particle Duality" and Ockham's Razor

In this paper, we apply Ockham's Razor [Encyclopedia Britannica, 1980] to interpret the general properties of the quantum physics database with regard to inferences of wave and particle behavior of quanta and with respect to the formulation of quantum mechanics. The term "quantum mechanics" specifically implies standard quantum mechanics (e.g., Messiah [1966]) and the acronym QM signifies the same. To provide a definite, though intuitively trivial, starting point, the Appendix defines the concepts of "particle" and "wave" in this context.

At the most basic theoretical and also popularly discussed level of quantum physics resides the concept of wave-particle duality. From the standpoint of research in physics, wave-particle duality is central to standard quantum mechanics; however, in terms of actual observation of quantum phenomena (addressed in the remainder of this paper), this concept is the source of significant confusion in both scientific e.g., Feynman et al. [1963, vol.1, p.37-6, para 6] (see below), and popular literature (see, e.g., Herbert [1985], p. 66; also see first sentence, https://en.wikipedia.org/wiki/Wave-particle_duality). Because "wave-particle duality" is ubiquitous in the literature and occupies such a high level in the theory, conceptual foundation, and history of quantum physics, scientists, in addition to technical writers, have defined the concept and interpreted standard observations similarly in the popular literature and physics texts. A popular overview like that of Herbert, therefore describes with sufficient detail the data on which the concept is based and the inferences that have been incorporated into quantum mechanics. For this reason, our discussion also occurs at a high level, in essence equally accessible to the public.

Our treatment begins with the demonstration, in Herbert's discussion (last paragraph, p. 66) and in almost every theoretical treatment of wave-particle duality, that the concept of an individual quantum acting like a wave, although ubiquitous, is an unverified inference from data. In fact, the data show that a wavelike pattern only appears within a very large number of events. No relevant data on the wavelike properties of a specific quantum exist. Rather, the "wave-particle duality," as often applied to a single quantum, is only a theoretical model and has not been observed directly.

To clarify this terminology, we employ William of Ockham's Razor [Encyclopedia Britannica, 1980], in which observation takes precedence over abstraction as an interpretive tool. According to Ockham, abstract terms are secondary, proceeding from observation. Among competing theories, the common statement of Ockham's Razor is to prefer or choose the simplest hypothesis or theory. As indicated in Section 3, Ockham's empiricism implies that descriptions of phenomena should incorporate the minimum necessary non-empirical information.

Ultimately, this allows one to separate and specify the particle and wave properties within quantum phenomena. Because scientists use words to visualize and explain the physical basis of observation and of theory, the elimination of imprecise terminology is a step forward in the application of quantum mechanics to engineering and design of future experiments and in proposing new projects for support.

The developments of this paper follow from observation of quantum phenomena. The remaining two subsections expand the discussion of particle and wave terminology as related to the Measurement Problem, for which the quantum Double-Slit Experiments (DSEs) serve as our paradigm to investigate particle and wave observations. These experiments date from 1803 when Young [1804] first demonstrated the classical wave nature of light via the classical interference pattern for light impinging on a double-slit assembly and thereafter illuminating a screen (also see Scheider [1986] or http://www.cavendishscience.org/phys/tyoung/tyoung.htm).

Section 2 provides a schematic of the experimental apparatus for classical and quantum DSEs and describes the empirical results. The section then addresses relevant features of standard theoretical interpretations of the data.

Section 3 briefly expands on Ockham's Razor, as originally enunciated and attributed, and also as related to more current scientific epistemology. Subsection 3.1 applies the original essence of Ockham's Razor to infer physical implications of particle and wave observations in the double-slit experiments. Subsection 3.2 addresses questions regarding the physical content of QM as related to or inferred from the DSE. In particular, the application of Ockham's Razor to the Double Slit Experiment leads to a new definition of a theoretical quantum system contrasting, for example, with a prevalent view that the wave function alone is sufficient for this purpose (e.g., Everett [1957]). Section 4 summarizes our conclusions on the meaning and implications of quantum phenomena identified with terms "particle" and "wave." The section then briefly considers conceptual paradoxes arising from the attempted use of single-quantum mechanisms in interpreting other types of quantum experiments, such as experimental tests of (a) Bell's Theorem [Bell, 1964] on locality and hidden variables in quantum physics and (b) the thought experiment of Einstein et al. [1935] as to completeness of quantum mechanics.

1.1. Standard Quantum Terminology

To provide a definite starting point, the Appendix follows Joos [1934] in defining the theoretical concepts of a particle and a simple wave. For classical wave phenomena in the DSE see, e.g., Marion [1968]. In some experiments, the individual quanta might be prepared with specific quantum numbers for properties like energy or polarization; our objective here does not require such complications. Beyond "particle" and "wave," one finds that related imprecise terminology can add considerable confusion to the quest for a more comprehensive formulation of quantum phenomena. The concept of "wave-particle duality" is particularly unclear and is our focus below. For example, one often reads that a quantum entity or "quantum" (terminology adopted here) may be described as "either [sic]" a particle or a wave (first sentence, https://en.wikipedia.org/wiki/Wave-particle_duality, and countless other sources). Feynman et al. [1963, 1965, vols.1, 3], for example, clearly agree with the picture of the particle and wave observations that we present below, yet in [1963, vol.1, pp.37-6, para 6] also muddy the picture for the reader by implicitly stating that "an electron behaves sometimes like a particle and sometimes a wave." However, we will find that the appropriate and relevant data (even as cited by Feynman et al.) do not exist to justify this latter description of an individual quantum as sometimes like a wave nor to determine how and when such single quantum behaviors apply in constructing particular mechanisms underlying data. As a result, explanations offered for the physics driving quantum double-slit interference run aground on paradoxes when unobserved single-quantum behavior is hypothesized (Section 2). William of Ockham offers an interpretation that is consistent with the data but which avoids such paradoxes (Sections 3-4).

1.2. The Measurement Problem

When broad terms like "particle" and "wave" are used to describe the implications of experiments, some details can become lost or obscured, actually leading to mathematical formalism which is inappropriate or questionable. To the point here, the standard terminology complicates the so-called "Measurement Problem (MP)," which is defined in various ways, but appears to have been formalized by von Neumann [1935, translated 1955]. In broad terms, the MP involves the interpretation of the quantum mechanical wave function and the wave function's relationship to experimental observations of quanta. This is intimately connected to the concept of wave function collapse (von Neumann [1955] process #1) and apparently originated with "wave function reduction" introduced by Heisenberg [1927]. While always cited, von Neumann's discussion of measurement is sufficiently opaque that Herbert [1985, page 146] switched the original designations of two types of quantum evolution processes identified by von Neumann. Everett [1957] provides an enlightening higher level summary of von Neumann's formulation, and DeWitt [Everett et al., 1973, pp. 169-176] provides a useful and easily followed discussion of measurement.

With respect to wave-particle duality and the double-slit experiment (DSE), the Measurement Problem regards the following facts (see, e.g., Herbert [1985], p. 66; the summary by Rolleigh [2010]; Venugopalan [2010]; Messiah [1966, e.g., Ch. 1]): (a) that an individual quantum is produced as a particle at the source and is always observed or detected as a particle at a particular location on a two-dimensional plane and (b) that the wave nature of quantum physics emerges in the spatial distribution of large numbers of quantum observation events (see next paragraph). When used, the term "wave function collapse" denotes an individual measurement event, as in the detection of a quantum particle in (a). Meanwhile, the (interference) pattern of (b) for large numbers of events is the only evidence of the so-called wave nature of quanta.

In other quantum experiments, a different spatial variable, e.g., scattering angle, parameterizes the distribution of detections, which, in the limit of large numbers of events, again approaches that observed for a classical wave. A well-known example is the observation of Bragg scattering of electrons as a function of scattering angle by Davisson and Germer [1928]; see also Venugopalan [2010].

The primary difficulty with the application of standard quantum mechanics (denoted herein as QM) to these experiments arises from attempting to attribute a wave nature to each individual quantum based only on the observation of wave patterns for large numbers of quanta produced under identical conditions. This bears repeating: the wave nature attributed to quanta actually appears only for a large number of events or quanta as a collective pattern, which approaches that of a classical wave! No direct observations have demonstrated that an individual quantum has a wave nature. This attribution to an individual quantum has become an unfounded assumption in many (almost all) discussions of mechanisms underlying QM observations, as in the double-slit experiments (e.g., Rolleigh [2010], Feynman et al. [1963, 1965], discussed in Sections 2-4 of this paper, and as described by Messiah [1966, Ch. 1] and Herbert [1985, pp.66-70]). Given the data which are available, along with the standard epistemology of science, and accepting the interpretive tool provided by Ockham, the author dispels this assumption in Section 2.

2. Double-Slit Experiments

The double-slit experiment (DSE) serves as a useful paradigm, for which an excellent summary is provided by Rolleigh [2010] and references therein. As examples, Tonomura et al. [1989] provide an experimental demonstration of the main elements of the DSE with an electron source and Rueckner and Peidel [2013] discuss the use of a discrete photon source. Marcella [2002] provides a quantum mechanical calculation of the interference pattern at the detecting screen of the DSE. Arve [2020] provides an up-to-date discussion of the standard interpretation of the QM wave function, which we will address in a publication following from our results below.

While eluding universally acceptable explanation at the single quantum level and thereby confounding the greatest physicists in history, the double-slit experiments are actually among the simplest. This is because the observations involve only the locations of the individual quanta on a detecting screen or two-dimensional array of sensors. We briefly mention other experiments in the discussion of Section 4.

2.1. Elements of DSEs

Figure 1 shows a basic schematic of Double Slit Experiments, following from ubiquitous examples, e.g., Feynman et al. [1963, vol.1, ch.37] and the colored and annotated redrawing of Feynman's figures by Gurappa [2019, Figure 1] in accord with the latter's quantum mechanical application. Our figure below is a redrawing of, but greatly different from, Gurappa's [2019] Figure 1, using some color symbols but eliminating other symbols and mathematical annotations that are specific to Gurappa's discussion. We have also aligned the schematic elements to reflect more accurately the horizontal symmetry axis of the apparatus.

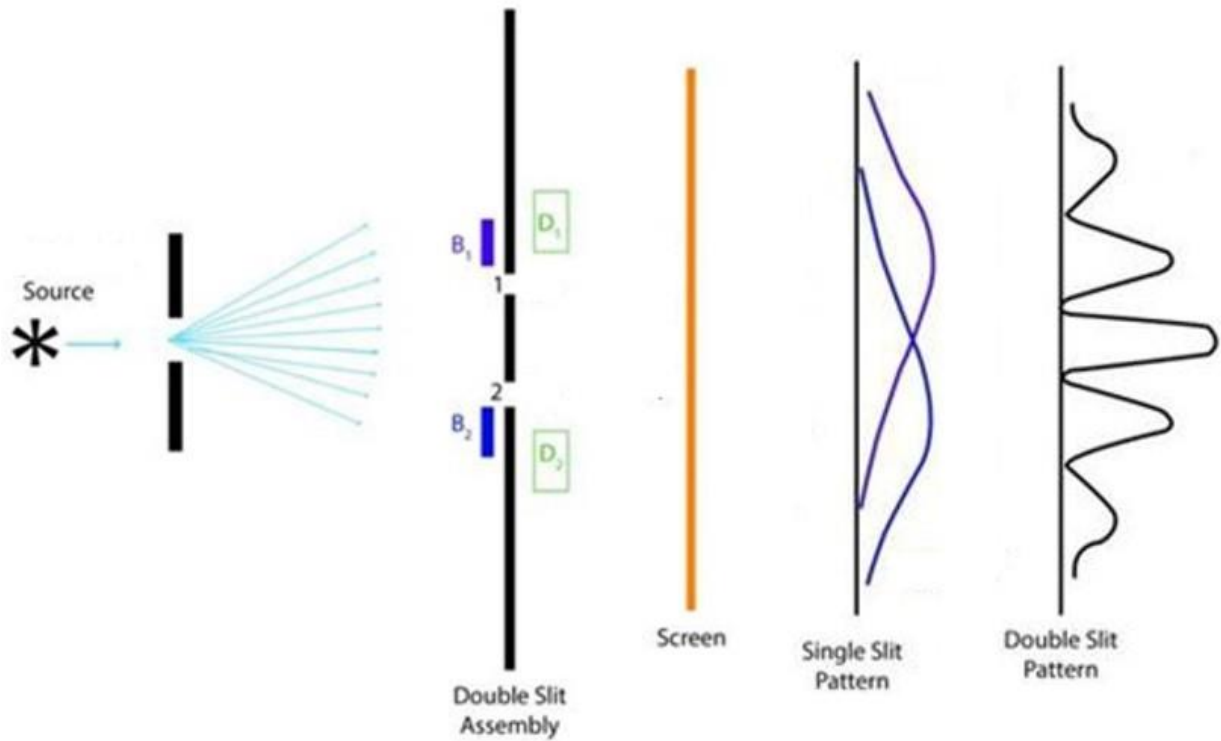

Figure 1 Elements of a Double Slit Experiment (e.g., see the similar schematic by Feynman [1963], Vol. 1, Figure 37-3).

The horizontal line of symmetry (not shown, but extending from left to right from the center of the "source" through the center of the interference peak at the far right) defines the positive z-axis. The positive x axis points perpendicularly upward from the z-axis in the plane of the drawing, and the positive y-axis points perpendicularly out of the page. The current view is downward toward the x-z plane along the negative y unit vector, $(-\mathbf{e}_y)$. A source on the left produces identical quantum particles (quanta) at a prescribed rate and traveling rightward, on average along the +z unit vector, $\mathbf{e}_z$, toward a barrier with two slits. In various experiments, one or the other slit might be blocked by $B_1$ or $B_2$; detectors $D_1$ and $D_2$ provide the option of determining through which slit a given quantum passes. The quantum particles strike the "screen," an array of detectors to form an x-y distribution or histogram of particle detections over time. If one or the other slit is blocked and following large numbers of individual detection events, then the histograms for the detector array will form a pattern that resembles the continuous diffraction "single slit" pattern due to a wave striking either the upper or lower single slit. The schematic superposes the two alternatives. If both slits are open, then after a large number of detection events, the x-y distribution of detections resemble an interference pattern like that observed by Young [1804] for light waves striking the slits. Variations of the experiment with regard to the combination of the types of devices {$B_i$} and {$D_i$} and their settings and the level of the source flux of quanta, demonstrate the apparent oddities of quantum physics with regard to the spatial distribution of detection events, e.g., Rolleigh [2010] (actual experiments), and [Feynman et al., 1963, vol. 1, ch. 3] (thought experiments).

In the classical picture, after the incident wave (e.g., light or sound waves) impinges on the slits, two waves emerge on the side of the barrier opposite the source and interfere before reaching the screen. In Rolleigh's example of the quantum experiment with a photon source, the slit width is 12 times the spatial scale (or wavelength derived from energy) associated with the emitted photons.

As indicated earlier, with respect to wave-particle duality, the Measurement Problem regards the fact that an individual quantum is always observed at the screen as a particle and by inference from studies of the source is produced as a particle at the source. Meanwhile, as explicitly pointed out, e.g., by Rolleigh, in the limit of large numbers of quantum observation events, the pattern or histogram of counts versus location on the screen approaches that of a continuous (classical) wave, first observed for light by Young in 1803. Again repeating: the wave nature attributed to quanta actually appears only for a large number of events or quanta as a collective interference pattern! No observations have demonstrated unequivocally that a particular individual quantum has a wave nature.

A conundrum of QM is that this process seems to occur for a very slow source emission rate, so that the individual quanta are emitted and detected one at a time. Over a long period of time, the collection of these successive events produces a two-dimensional histogram of counts apparently identical to the distribution that occurs when the source emits a large flux of quanta. In the high flux case, one might conjecture that the quanta passing through the pair of slits could interact with each other after passage, which Rolleigh [2010, Section 4] dispels by appealing to the experiments with separated, successive, single quanta.  The bottom line is that no one knows for sure what happens because no coincident observations of the transition between emission and detection of a given quantum are available. Ultimately, as the number of emitted quanta and observation events increases, the interference pattern demonstrated for a classical wave by Young emerges in the two-dimensional histogram of quantum counts.

For the very slow emission rate experiment, the entertaining mystery regards what happened after emission of separate (successive) particles at the source and prior to detection at the screen to produce the observed interference pattern over large numbers or events. Proposals of single-quantum mechanisms abound (e.g., Aharonov et al. [2017]; Rolleigh [2010]). As an example, Rolleigh mentions and then dispels an imaginary process in which an individual quantum interacts with both slits during transit. How would a whole particle do that? Another frequent proposal is that, unobserved, the quantum entity, e.g., electron or photon, acts as a wave, interfering with itself, and then becomes a particle or is somehow detected as one at the screen. Again, to everyone's disappointment, for a specific element of the set of events that produced the pattern on the screen, no coincident data are available to reveal what happened to a specific individual quantum during transit. For this reason, the conundrum passes from the realm of science to the realms of mathematics or philosophy, at best, or of fantasy at worst.

Here, the present author addresses the issue of the transit region by applying Ockham's philosophy of minimal abstraction or of invoking the simplest underlying hypothesis. A more modern meaning is to add the least amount of non-empirical (i.e., a priori theoretical) information to the actual data (pure empirical information), in order to retrieve the most straightforward inference (a posteriori information, see e.g., Tarantola and Valette [1982]) consistent with the Correspondence Principle.

The choice of minimal non-empirical information plus the constraint of consistency with classical (i.e., macroscopic) phenomena and the Correspondence Principle, separates our results from other treatments and interpretations that involve hypothetical single-quantum behavior. The latter approaches invariably introduce paradoxes contradicting fully-validated classical theories (e.g., special relativity) and concepts (e.g., causality, locality), both categories of which are also consistent with common sense based on observation of the macroscopic world. The paradoxical single-quantum descriptions therefore disagree with the essence of the Correspondence Principle, which is fundamental to quantum theory (see next paragraph and Sections 3 and 4).

A well-known example is nonlocality (e.g., Aharonov et al. [2017], Bell [1964]; see Section 4), which has not yet been directly observed for a single quantum in a physical or measurement process. Instead the basis for asserting "nonlocality" at the single quantum level derives from Bell's theoretical model of locality, which was assumed, but not proven, to be exhaustive (see Section 4 on Bell's Theorem) and which was used to define hidden variable theories for testing against a statistical quantum data set. The hidden variables theories defined by Bell failed the test of reproducing the result of standard quantum mechanics for statistical data. That was the point of the paper. More importantly to us, this statistical result does not directly address the nonlocality of interactions for a particular individual quantum (in fact, see Section 4.1) as Bell had already stated [Bell, 1964, Section II, paragraph 1, last sentence]:

"… the initial quantum mechanical wave function does *not* [sic] determine the result of an individual measurement …"

Remarkably, in spite of this shortcoming, Bell (Section 4.1 below) continued on to mention quantum nonlocality of individual quanta or in individual measurement events, which has since become an accepted theoretical construct for discussing quantum phenomena and interpreting experiments (e.g., Aharonov et al. [2017]).

Such proposals derive in DSEs from the lack of concomitant data on both (a) transit of an individual quantum across the region of the slits and to the screen and (b) the recording of that quantum's arrival at the screen. The Measurement Problem is therefore a problem for which quantum mechanics appears to fall short of providing a unique model. Epistemologically, the scientist requires data to clarify the underlying physics. Causality, the Correspondence Principle, and common sense have demonstrably suffered in theoretical development at the quantum level because appropriate data do not exist.

2.2. Observations and Issues to Be Addressed Via Ockham's Razor

The empirical approach of William of Ockham (Ockham's Razor) provides needed clarification of quantum theoretical concepts via observations of the quantum DSE. We infer that a basic (well-known) principle of quantum mechanics (the Correspondence Principle) actually provides the closure required of a "complete" theory (see Section 4). As argued below, the Correspondence Principle of QM and Ockham's Razor allow us to calculate the results of all existing DSEs regarding wave and particle properties of quanta while successfully addressing the elements of the associated measurements:

(1) Individual quanta are emitted and observed as whole particles.

(2) Large numbers of identical observations approach the interference pattern produced by a continuous classical wave or quantum-mechanical wave function.

Section 3 applies Ockham Razor to infer physical implications of observations (1) and (2) in the double-slit experiments. Subsections address follow-on questions regarding the physical content of QM as related to the DSE. The questions consider properties often attributed to an individual quantum in theoretical mechanisms hypothesized to govern the time period between emission and detection of each quantum, whether emitted sequentially or within a near-simultaneously emitted collection of individual quanta. Since these questions are intimately tied to primary observations (1) and (2) above, we group them under the next sequential index:

(3) Should quantum mechanics provide a single-particle mechanism for observations of both (1) and (2) above? For example,

(3a) Do individual quanta have wave properties in the course of a DSE measurement?

(3b) If (3a) is were to be answered, "No," then how should wave properties be imposed on a realistic model of an individual quantum so that a collection (of large numbers of quanta) forms the pattern of (2)?

As indicated earlier, Section 4 briefly comments on terminology that might be clarified or at least restricted by our interpretation of (1) and (2) and Ockham-inferred conclusions regarding question (3). There we consider the potentially misleading textual descriptions that can accompany reports or analysis of quantum phenomenology inferred from experiments, including the gedanken experiment of Einstein et al. [1935], as to completeness of quantum mechanics.

3. Ockham's Razor and Interpretation of DSEs

The primary tool used here for interpreting the DSE and associated theoretical treatments is William of Ockham's oft-attributed razor, "Entities are not to be multiplied beyond necessity." [Encyclopedia Britannica, 1980]. Modern statements include, "The simplest explanation is most likely the right one" [https://en.wikipedia.org/wiki/Occam's_razor].

The latter view takes liberty with history. Ockham was generally not amenable to theory or more specifically, abstraction, instead favoring a straightforward empirical statement [Tsanoff, 1964, pp. 217-219]. In this regard, Ockham conflicted with current scientific epistemology, which involves the constant comparison of current mathematical theory with experiment and the resulting upgrade of theory and the development of follow-on experiments.

On the other hand, scientists and popularizers of science often develop a dialog which seeks to explain the implications of the mathematics without delving deeply into the mathematical and experimental details. This can lead to misconceptions among scientists and engineers, which can obscure the experimental facts. Here Ockham shines in emphasizing the empirical facts over textual descriptions. Adding a textual description of possible causes or of a plausible mathematical theory of underlying mechanisms can add erroneous or unnecessary external information to analysis and interpretation of the data. This can obscure the fact that data constitute all of our actual knowledge of physical systems, i.e., facts. From the standpoint of fidelity of information content, Ockham's Razor is actually optimal in adhering only (or possibly, almost entirely) to data, which constitute the maximum verifiable information available, while minimizing non-empirical sources of error.

To quote Ockham via Tsanoff [1964]: "Universals do not exist in nature; to believe otherwise is 'simply false and absurd.'" "For example, remarkably apropos of the DSE observations, Ockham rejected the concept of "motion" in favor of "the reappearance of a thing at a different place" [Encyclopedia Britannica, 1980]. Thus, he considered interpretation of data to reflect as directly as possible the actual observations of "things." A modern exposition of Ockham's approach might allow the addition of some, but minimal, abstraction. Consistent with this, the present discussion allows the possibility of moving closer to a more modern viewpoint that nevertheless uses the fewest possible theoretical assumptions to describe or organize experimental data.

However, we will not do so at the expense of introducing universal microscopic concepts that are counter to what we readily observe at macroscopic scales. Examples of verboten characteristics include violations of causality or of the constraints of general relativity on information transfer; both violations contradict the Correspondence Principle, which in a general sense requires agreement with macroscopic (i.e., classical scale) observations. Nevertheless, as Ockham would have insisted, the data are primary, and moving beyond observation can lead to meaningless mental gymnastics at best, or to fantasy at worst.

3.1. Implications of Ockham's Razor for Interpreting DSEs

3.1.1. Implication 1: Individual Quanta Are (Whole) Particles Only

Indeed, the first and only observation of a given quantum particle by the double-slit experiment is a count registered at a particular position on the detection screen. Based on studies of the source, Ockham would infer that the detected particle was emitted as a particle earlier at the source. Based on the observation at the screen and this inference on the source, Ockham's razor infers that the quantum is a whole particle, and only a whole particle, throughout its transit from emission to detection.

Note that the position measured on the screen (i.e., the particular detector) is a macroscopic observable and is therefore classical. This is not surprising since a particle is in its essence a classical construct [Joos, 1934]. We adopt the view of the simplest interpretation, as provided by Ockham, that the observed quantum is a particle throughout the process, from emission, during transit, and to detection at the screen. The data are consistent with this. Hence the (ultimately classical) observed particle location satisfies the Correspondence Principle [e.g., Messiah, 1966, pp 7-11, 29] trivially: The location of the observed quantum particle in the classical limit (i.e., at macroscopic scales here) is a visible registration or indication (e.g., a blip) at a particular detector with macroscopic accuracy, as defined by finite detector size as a fraction of screen size.

Further, the empirical approach of Ockham infers that observed individual quanta are whole particles only. That is, an individual quantum does not split as a wave could [e.g., Rolleigh, 2010]. Therefore, Ockham infers, for example, that an electron emitted at the source and then counted at the "detector screen" of a two-slit experiment must be a whole particle at all points in between. No quantum experiment contradicts this "simplest" picture, and therefore, one finds no empirical support for the Copenhagen or von Neumann proposal of wave function collapse at observation.

3.1.2. Implication 2: The Quantum Wave: A Collective Property

The so-called wave nature emerges in an experiment only in the limit of large numbers of particle observation events. In that limit, the two-dimensional histogram of recorded counts approaches the classical two-dimensional intensity distribution of light measured by Young [1804] at the screen. The standard QM wave function (i.e., squared modulus) thus serves only as an idealization of the wavelike distribution of counts at the screen, which is the only quantum observable other than the original counts appearing sequentially, depending on the source emission rate.

Thus, the quantum wave in the DSE in the limit of large numbers of events is simply a demonstration of the Correspondence Principle (CP). A key element is that the array of counts is macroscopic, and therefore observation of the emerging wave is also at macroscopic or classical scale. As a result, given the data for a classical wave, the Correspondence Principle of quantum mechanics is sufficient to predict the wave observed as a collective property of the quanta. The wave occurs because the CP decrees that such a wave is a necessary condition of quantum phenomena. Admittedly, the author cannot know what Ockham would have thought of the Correspondence Principle. However, the CP is an axiom imposing a directly observable constraint on the interpretation of direct macroscopic observations of the quantum DSE, which would appear to satisfy the desire for minimum abstraction.

Meanwhile, Ockham tells us that reality consists of observables and that quantum mechanics is "complete" based on all extant data (compare Einstein in Section 4). To make progress, one must find probes of the intervening transit region following emission for each specific quantum that is detected at the screen.

3.2. Theoretical Implications Regarding Physical Content of Quantum Mechanics

3.2.1. Definition of a Quantum System

Based on the application of Ockham's Razor to DSEs, we conclude that a theoretical quantum system consists of at least one elementary quantum particle plus a wave function specifying the statistical distribution and properties of a large number of such quantum particles. More generally, one might infer that a quantum system could be defined by a "unit set" of one or more quantum particles plus the appropriate wave function describing an ensemble of such unit sets. For DSEs involving a source of composite quantum objects, the unit set would consist of the particles constituting an individual composite quantum object, and the wave function would then describe the statistical results of detections of an ensemble of identical such unit sets. In the simplest DSEs, as discussed here, an elementary quantum particle (e.g., electron or photon) emitted at the source would define the unit set. The wave function then corresponds to the emerging histogram of particle detections at the screen or detector array in the limit of large numbers of observations.

3.2.2. The inferred Answers to Question (3) of Section 2.2.

Ockham worked only from observations to describe physical processes; a modern restatement of the principle allows a minimal set of axioms to construct a theoretical description with which to predict experimental results. Ockham would have warned that the latter abstraction might lead to conceptual problems as indicated in Sections 2 and 4. Meanwhile, we close the loop on Questions (3a) -3(b) under Question (3) Section 2.2, repeated here:

(3) Should quantum mechanics provide a single-quantum mechanism for observations of both (1) the particle nature and (2) the wave nature of quanta? Based on Ockham's Razor and empiricism, and as pointed out by Bell [1964, Section II, paragraph 1, last sentence] (quoted in the last paragraph of Section 2.1), the answer is no. For example,

(3a): Do individual quanta, taken separately, each have wave properties in the course of a DSE measurement?

To date, the data on the double-slit experiment fully support the simplest inference, following Ockham's Razor, that an individual quantum is a whole particle from emission through transit to detection at the detector screen. The available data from DSEs and other experiments over the last century show only the existence of individual quanta as particles. No observation or experiment has verified the hypothesized wavelike character of any particular individual quantum. Rather one observes that the collection of large numbers of such quanta can form a wave like that observed classically. Assuming the viewpoint of Ockham, one must answer question 3(a) above in the negative, i.e., "No."

(3b): Since (3a) is answered, "No," how should wave properties be imposed on a realistic model of an individual quantum so that a collection (of large numbers of quanta) forms the interference pattern of observation (2) in Section 2.2?

This question contains an implicit assumption that data imply that wave properties must necessarily be imposed on a realistic model of an individual quantum. The data do not demonstrate this and are equivocal; the question is therefore inappropriate. Certainly a natural question regards what happens between emission and detection of an individual particle, such that large numbers of quantum particles in the DSE are constrained to approach a classical wavelike spatial pattern. As mentioned in Section 1, paragraph 2, the required data on specific individual quanta (i.e., the same quantum during transit and at the screen) do not exist.

The historical assertion that only one measurement can be made on a quantum, thereby altering the quantum state (here, the properties of the quantum), seems to obviate such data. For the DSE, the so-called quantum wave nature is observed to be a collective phenomenon involving spatial coordinates, specifically of large numbers of quanta detected individually at the screen. Ockham would say that the lack of data demonstrating the wave nature of a single quantum automatically obviates such an attribute from the physical content of quantum mechanics. Consequently, according to Ockham's reasoning, an attempt to use quantum mechanics to construct such a model of a single quantum is irrelevant to the quantum physics database, which is already addressed successfully by quantum mechanics. Remarkably, Bell [1964, Section II, paragraph 1, last sentence] agrees about the limitations of quantum mechanics, as quoted in Section 2.1. Further a wavelike single-quantum model could contradict quantum mechanics and therefore could lead to disagreement with or inapplicability to data or could lead to paradoxes or contradictions with physical principles like causality or locality, as happens (Section 4). The latter would violate the thrust of the Correspondence Principle. The physics world cannot have its cake and eat it, too.

4. Summary and Application to Interpreting Other Types of Experiments

Here the author should at least mention the implications beyond double-slit experiments and others involving spatial coordinates only. The particle nature of individual quanta appears naturally at the event of observation by a detector. The focus of quantum mechanics is the wave nature of the observations. The above application of Ockham's Razor clarifies the observation of the wave nature as occurring at macroscopic scales in the limit of large numbers of events. Following Ockham, the Correspondence Principle achieves primary importance as a fundamental principle of QM, in being sufficient to predict the quantum wave nature from classical wave DSEs. Most importantly, this prediction of quantum mechanics addresses a macroscopic and collective property of (large numbers of) individual quanta, taken as a whole. As a result, no observational basis exists for applying the quantum wave function to individual particle dynamics (Bell [1964] agrees, as quoted at the end of Section 2.1).

Conversely, following from Ockham's Razor, single-quantum mechanisms and behavior and dynamics of specific quanta are not observed directly in quantum physics, and theoretical discussions postulating single-quantum phenomena like self-interference have no place in analyzing what are necessarily statistical data from experiments on quantum systems. While mathematically interesting and entertaining, the physical relevance of the wave observations to the study of individual quantum phenomena within the transition region between emission and detection is lacking and can lead to fundamental misconceptions. Other systems and experiments can involve (a) additional variables, e.g., including quantum numbers, attributes of individual quanta, like charge, angular momentum, or polarization, or (b) interactions of quanta, producing correlations, for example. The wave function in such cases must be more complex, at least in terms of information content. The general wave function therefore can involve a superposition of quantum states, each of which can be represented by one of a set of orthogonal wave functions (e.g., Messiah [1966], Everett [1957]). Further, the analogs in the classical limit for direct comparison with bound or entangled quantum systems are not readily apparent (to the author at least).

Even so, as asserted, e.g., by Rolleigh [2010], and demonstrated by standard quantum mechanics texts, e.g., Messiah, [1966], observations of such quantum systems consist of large numbers of individual quanta, e.g., produced by state transitions or emanating from the region in which entanglement took place. Based on the interpretation of the DSE via Ockham's Razor and the Correspondence Principle, inferences about the wave nature of such a system or phenomenon derive from the collective properties of an observed set of events. Therefore, by the CP, wave-function-based calculations of quantum properties (variables), in the form of quantum mechanical expectation values of the corresponding operators, must agree with and correspond to averages of the respective measurements for large numbers of individual quanta.

4.1. Example: Bell's Theorem -- Hidden Variables and Nonlocality

Bell's Theorem [Bell, 1964] computes from the wave function an expectation value for the angular correlation of vector components (e.g., polarization vector components or quantum particle spin angular momenta) in selected directions, denoted **a** and **b** by Bell in the case of two particles with spin operators $\sigma_1$ and $\sigma_2$. For the case of a singlet state for two spin-1/2 particles, Bell computed the expectation value to be

$$\langle (\sigma_1 \cdot \mathbf{a})(\sigma_2 \cdot \mathbf{b}) \rangle = - \mathbf{a} \cdot \mathbf{b} \quad (1)$$

Reiterating our discovery above for the DSE, the quantum wave nature expressed in Eq. (1) is a collective phenomenon as described by the Correspondence Principle in the limit of large numbers of individual particle events. In this case, the theoretical correlation function following from the quantum wave function must estimate the correlation as an average over a large number of observed pairs of simultaneously emitted (entangled) quantum particles for each instrumental setting, **a** · **b**, which selects the angular separation of the measured components (see for example the simple discussion by Herbert [1985] or Mermin[1985]).

As in the case of individual particles in the DSE, data on locality or nonlocality of the combined measurement process for each (separate) observation of twin particles are not available. As quoted in Section 2.1, Bell realizes that quantum mechanics does not apply to individual events [Bell, 1964, Section II, para. 1, last line]. Rather, Bell defines a mathematical model of the correlation measurement in terms of local hidden variables, from which he then computes a theoretical expectation value for a statistical ensemble of observation (measurement) events. Statistical experiments on the problem posed by Bell have verified that the hidden variable theories, as defined by Bell's model, are incorrect and that standard QM gives the correct correlation. For this reason, Ockham's Razor would select standard quantum mechanics over the hidden variables theory defined by Bell, and the story would end here.

Curiously, Bell [1964] continued on to outline a hidden variables theory augmented with nonlocality. With regard to entangled pairs of quanta which propagate away from each other to be measured by separate identical devices:

"In a theory in which parameters are added to quantum mechanics to determine the results of individual measurements, without changing the statistical predictions, there must be a mechanism whereby the setting of one measuring device can influence the reading of another instrument, however remote. Moreover, the signal involved must propagate instantaneously, so that such a theory could not be Lorentz invariant."

Ockham's Razor would naturally have selected standard quantum mechanics, which is the simpler alternative and which already agrees with the extant data. Further, no observations of nonlocality in a specific quantum event are available. Also, Bell immediately recognizes that nonlocality in individual events violates classical causality, as defined by the limitations of special relativity, and therefore is counter to the essence of the Correspondence Principle.

As cautioned in earlier sections, Bell's hypothetical comment has spawned a confusing landscape of unverified assumptions about nonlocality of quantum phenomena. Although Bell is specifically proposing nonlocality to rectify hidden variable theories as he defined them, the concept of nonlocality, once released from Pandora's Box, has entered the lore of quantum phenomena [Herbert, 1985, pp. 215-226]. More troubling, nonlocality has become a component of serious theoretical models even for individual quanta, e.g., as invoked recently by Aharonov et al. [2017], and many have concluded that quantum interactions might be "nonlocal."

For these reasons, the author steps outside of our final conclusion based on Ockham's Razor and empiricism to discuss Bell's paper further with regard both to hidden variables and to quantum nonlocality as a standalone property. Implicit in Bell's interpretation is the assumption that his model of hidden variables exhausts all possibilities and most importantly, that the apparent implication that he states for hidden variables models of a given individual event is correct. As Bell himself found with Von Neumann's treatment of hidden variables theories, such models can fail to cover all possibilities. Similarly, Bell's proposed single-event mechanism for fixing hidden variables theories is not demonstrably the only possible physical basis for the quantum mechanical result. In fairness, he did not say so, but as we have stated, interpretations by others of textual descriptions like the quote above have been misleading.

To repeat, standard popular descriptions of Bell's Theorem assert that quantum mechanics and quantum phenomena are nonlocal (see Herbert [1985], pp. 215-226]). With regard to Ockham's empiricism, the concept of quantum nonlocality flies in the face of what we observe in the macroscopic world. This concept therefore violates the essence of the Correspondence Principle, which is fundamental to quantum mechanics. On the other hand, Bell's "model" of locality might not be provably exhaustive, thereby nullifying or weakening significantly his arguments even before his inclusion of nonlocality.

Fortunately, physics as a discipline requires actual direct observation to verify a hypothetical mechanism. Because Bell's related single quantum event nonlocality mechanism is neither observed nor verified directly, the author follows Ockham to reject single event nonlocality as a necessary element of the optimal theory, and instead to favor standard quantum mechanics as the preferred theoretical description of quantum ensembles, i.e., in the limit of large numbers.

This situation points out the shortcoming of our quantum database: actual data on the interaction of specific entangled pairs of quanta with each other and with instruments are lacking. Even without worrying about what possibility Bell might have missed in his ansatz on hidden variables theories, Ockham would therefore reject the assertion of "nonlocality" to describe the implication of such statistical experiments. Instead, Ockham would insist that such inferences cannot be made until actual, relevant, empirical demonstrations of occurrences in individual events.

Finally, the quantum mechanics (theoretical) result, Equation (1), agrees with experiment, thereby satisfying Ockham's requirement for the validity of standard quantum mechanics. For these reasons, Bell's complicity in this confusion over quantum nonlocality is moot. As indicated above, Ockham's Razor rules out the superfluous concept of quantum nonlocality, which has not been observed at the microscopic level, and which is counter to classical (macroscopic) observations, therefore violating the essence of the Correspondence Principle.

4.2. Einstein and Completeness of Quantum Mechanics

Einstein et al. [1935] defined the concept of a "complete" physical theory to mean that "every element of the physical reality must have a counterpart in the physical theory." However, that statement is a more subtle and sophisticated approach to explaining his reservations about the indeterminacy of quantum mechanics -- often expressed as randomness or a statistical nature in his more well-known comments. The 1935 paper considered entangled particles, likely inspiring Bell's paper (e.g., Herbert [1985], p.212).

The sticking point is the specification of "reality." Einstein implied that he would be the judge of what constituted "physical reality." Based on his example in 1935, we infer that quantum reality to Einstein would include pairs of complementary variables that are simultaneously observable classically for a specific object. He chose position and momentum of a given quantum. At quantum scales, position and momentum of a quantum object are not measureable simultaneously, which Einstein et al. recognized to be a shortcoming of their gedanken experiment on an entangled system of two particles (see Einstein et al. [1935], next-to-last paragraph).

Ockham would likely have weighed in differently on the terms "reality" and "complete." To paraphrase his point of view, a phenomenon or property is real if and only if observable, as represented in experiments by data. Precise, simultaneous values of a pair of complementary quantum variables for a given quantum object are not observable by compatible experimental arrangements [Messiah, Ch. IV, para. 17, p. 153]. Therefore, such data would not exist within the extant quantum database. To Ockham, if a theory describes all extant data, as quantum mechanics does, then that theory provides a complete description of reality. Note that the data to which quantum mechanics applies are statistical, applying to large numbers of particular observations of identical objects. A careful reading of Einstein et al. [1935] shows in fact that the discussion speaks of "systems" and "particles" interchangeably and that this picture of a quantum particle underlies the discussion. At the same time, the authors recognize that quantum mechanics represents a quantum state as a wave function, which applies only to the properties of a statistical ensemble of identical particle observations.

Meanwhile, a necessary condition of the scientific method is that a fully successful theory must predict the results of all proposed measurements. This is where Ockham's view might depart from the epistemology of science, since for him, prospective experiments would not qualify as real.

By his definition of reality and, therefore, contrary to Einstein, Ockham tells us that quantum mechanics is "complete" based on all extant data, which are statistical in nature. Quantum physics on the other hand includes both the theory (QM) and the quantum physics database. Instead of QM, Ockham would insist that, if any element of quantum physics is to be considered incomplete, then the database must be that element. An example would be the unobserved transit region for a specific quantum within the double-slit experiments,. To make progress on a single-quantum mechanism underlying the DSE interference pattern, says Ockham, one must observe a specific quantum particle at emission, detection, and in transit. The longstanding view of quantum phenomena is that this is impossible because observation will change the state of the quantum [Einstein et al., 1935, paragraph 8].

Generalization of the results of this section and applications of Ockham's logic to other problems or to various formulations of quantum mechanics, e.g., Everettian interpretations or Bohmian mechanics (e.g., see Herbert, [1985]) appear straightforward and are therefore tempting. However, such a survey is outside our objectives in this paper and will await a future publication.

Acknowledgements

The author acknowledges support by the United States Civil Service Retirement System and affiliation with, and computer and library support by, the Naval Research Laboratory (NRL) Voluntary Emeritus Program. This has made available interesting problems and deep physical insights of the outstanding scientists of the Geospace Science & Technology Branch of the Space Science Division. The author gratefully acknowledges the generous permission of N. Gurappa to use Figure 1 of his article, cited below, to obtain symbolic elements for constructing Figure 1 of the present article. This is an unusual article for the author, in terms of the reach to such a fundamental level. NRL is not responsible for opinions expressed herein by the author or controversy that the article might cause.

Appendix: Definitions of "Particle" and "Wave"

A particle [Joos, Chap. 5, p.75] is an object whose extension in space may be neglected, and which, therefore, may represented as a point in spacetime, endowed with definite mass and charge (here updated to include the possibility of other detectable parameters). The practical manifestation would be a quantum object that could only trigger at most a single counter or detector of the smallest achievable dimensions. In practical terms, this means that the size of the object lies below the spatial discrimination of our current technology. From the standpoint of size and endowment with mass, charge, or another detectable parameter, quantum particles are, therefore, indistinguishable from their classical counterparts, i.e., classical particles. One nuance that we do not address are microscopic objects, e.g., atoms, which can be imaged or detected with proper techniques as structureless bumps, i.e., with discrimination of size (e.g., Teper e al. [2006]). However, accomplishing this discrimination for a DSE detector screen seems to be beyond the current capability.

As defined in early physics courses, a "wave" is a traveling disturbance for which upper and lower limits on spatial extent or spatial variation can be measured. In practical terms, a wave is finite, the size being measurable by laboratory equipment, and is therefore macroscopic, in contrast to a particle. As an example, the simplest general form of a wave in one dimension is [Joos, Chap. 2, Sec. 5]

$$u(x, t) = F(x - vt) + G(x + vt) , \quad (A.1)$$

where x is position, t is time, and v is a phase velocity that is related to measureable wavelength $\lambda$ and frequency f by

$$v = \lambda f . \quad (A.2)$$

The function $F(\cdot)$ represents propagation in the positive x direction from the origin and $G(\cdot)$ represents propagation in the negative x direction. This quickly becomes more complicated with higher spatial dimension and a background medium.

Our interests are (1) individual quanta as separate particles and (2) the resulting wave, which defines a spatial distribution of individual quanta in the limit of large numbers of detection events (counts). That is, the direct observation is of each quantum object or "quantum." For a given source of quanta, as the number of detections increases, the quantum mechanical "wave" emerges in the spatial pattern of quantum detections via a similarity to the spatial pattern of a continuous function observed in the identical experiment with a continuum (classical) wave source. As discussed in the present paper, the spatial distribution of quanta relates directly to the quantum mechanical (QM) wave function. Outside our scope are more complex experiments measuring statistical properties of a statistical ensemble of individual quanta in a complex state, e.g., as in the statistical correlation of spins of coincident pairs of quanta observed at different detectors. [Bell, 1964].